\documentclass{article}
\usepackage{blindtext}
\usepackage{dirtytalk}
\usepackage{geometry}
\usepackage{authblk}
\usepackage{amsfonts,amsmath,amssymb,amsthm}
\usepackage{graphicx}
\graphicspath{ {./downloads/} }
\usepackage{caption}
\usepackage{float}
\usepackage{dirtytalk}
\usepackage{subcaption}
\title{Metriplectic Heavy Top: An Example of Geometrical Dissipation}
\author{ By Michael Updike\\ Supervised by Dr. P. J. Morrison
}
\affil{Undergraduate Thesis \\ Department of Physics, The University of Texas at Austin}
\usepackage{biblatex}
\date{December 2022}
\begin{document}
\maketitle
\renewbibmacro{in:}{}
\section*{Abstract}
 Recently, Morrison and Updike \cite{mine} showed that many dissipative systems are naturally described as possessing a Riemann curvature-like bracket, which similar to the Poisson bracket, generates the dissipative equations of motion once suitable generators are chosen. In this paper, we use geometry to construct and explore the dynamics of these new brackets. Specifically, we consider the dynamics of a heavy top with dissipation imposed by a Euclidian contravariant curvature. We find that the equations of motion, despite their rather formal motivation, naturally generalize the energy-conserving dissipation considered by Matterasi and Morrison \cite{metfrb}. In particular, with suitable initial conditions, we find that the geometrically motivated equations of motion cause the top to relax to rotation about a principal axis. 
 \section*{Definitions and Conventions}

 All repeated indices are assumed to be summed over unless otherwise specified.
 We denote the phase space manifold of a system $\mathcal{Z}$. We use $z^i$ to denote the coordinates of $\mathcal{Z}$. The symbols $f,g,h, F,G,K,$ and $N$ are always used to represent smooth ($C^\infty(\mathcal{Z})$) functions.
 The tangent bundle of $\mathcal{Z}$ is denoted $T\mathcal{Z}$, and the cotangent bundle $T^*\mathcal{Z}$. 
 
 Here, $\alpha, \beta,$ and $\gamma$ always represent one-forms.  The space of one-forms is denoted $\Omega^1(\mathcal{Z})$.  The symbol $\mathbf{d}$ represents the exterior derivative, which acts on a function $f(z)$ as 
 \[
 \mathbf{d}f = \frac{\partial f}{\partial z^i}\mathbf{d}{z^i}.
 \]
 The exterior derivative of a coordinate function is strictly formal. In coordinates, a general one-form may be written 
 \[
\alpha = f_i(z) \;\mathbf{d}z^i.
 \]
 A one-form $\alpha$ is said to be exact if $\alpha = \mathbf{d}f$ for some $f$. 
 
Given a tangent vector $v$, we use $v[f]$ to mean $v$ acting on $f$. We use $\left<\beta, v \right> $ to denote the canonical pairing between a vector and a one-form. 

We use the word bracket to mean a smooth map from some number of smooth functions to a single smooth function. We demand that a bracket is both linear and a derivation in each of its arguments. In particular, we 
use $\{\cdot ,\cdot \}\colon $  $ C^\infty (\mathcal{Z})  \times  C^\infty (\mathcal{Z}) \to C^\infty(\mathcal{Z})$ to denote the Poisson bracket. 
The Poisson bracket gives a bivector field
\[
J^{ij} = J(\mathbf{d}z^i,\mathbf{d}z^j) \colon = \{z^i, z^j \} 
\]
and also an anchor map $J^\#: T^*\mathcal{Z} \to T\mathcal{Z}$ defined implicitly by 
\[
\left<\beta, J^\#(\alpha) \right> \colon = J(\alpha, \beta). 
\]

We use $D$ to represent a contravariant connection, also called a contravariant derivative, which gives the \say{derivative} of a-one form with respect to another one-form. A contravariant derivative is similar to but not the same as  a covariant derivative.
 \section*{Introduction}

Given a degenerate Poisson bracket $\{ \cdot, \cdot \}$, there exist distinguished functions $S$ called Casimirs that Poisson commute with all functions $f$ 
\begin{align}
\{f, S\} = 0.
\end{align} 
Casimirs represent, potentially, the entropy of a system. Given that the dynamics of \textbf{any} Hamiltonian system with the Poisson bracket $\{\cdot, \cdot \} $ is constrained to surfaces of constant $S$, a dissipative structure is required to create dynamics that respect the second law of thermodynamics. First developed by Morrison and others, metriplectic dynamics is a systematic way to add energy-conserving dissipation to an otherwise Hamiltonian system (cf.\cite{para}). 

Suppose we have a Hamiltonian system with Hamiltonian $H$ and a Casimir $S$. We consider a symmetric bracket $(\cdot, \cdot)$ such that, for all functions $f$, 
\begin{align}
(f,H) = 0.
\end{align}
The metriplectic equations of motion that both preserve energy and increase $S$ are 
\begin{align}
\dot{f} = \{f,H \} + (f,S).
\end{align}

In this paper, we first use the framework of both Riemannian and Poisson geometry to rephrase metriplectic dynamics as a geometrical theory. We then use this new approach to metriplectic dynamics to recreate a bracket first introduced by \cite{original} and used in \cite{metfrb} for control of a rigid body. Afterward, we naturally generalize the construction to the heavy top system, constructing a family of dissipative theories which we collectively call the \say{metriplectic heavy top.} Finally, we simulate the dynamics of a handful of these theories, finding asymptotic relaxation of the heavy top. 
\section*{Riemann-Poisson Geometry}
Suppose we have system described on some Poisson manifold $(\mathcal{Z}, \{ \cdot, \cdot \})$ with coordinates $z^i$. The Poisson bracket on functions is naturally extended to $T^*\mathcal{Z}$ by the Koszul bracket $[\cdot,\cdot]$, which for exact forms reads \cite{holonomy}\cite{rpliegroups}
\begin{align}
[\mathbf{d}z^i, \mathbf{d}z^j] \colon = \mathbf{d}\{z^i, z^j\}. 
\end{align} 
The Kozul bracket gives us a natural Lie bracket on forms.
Given a (pseudo-)metric 
\begin{align}
g^{ij} = g(\mathbf{d}z^i, \mathbf{d}z^j)
\end{align}
we can define a contravariant Levi-Civita connection $D$ on $\mathcal{Z}$ via the formula
\begin{align}
2g(D_{\alpha} \beta, \gamma) = & J^\#(\alpha)[g(\beta, \gamma)] - J^\#(\gamma)[g(\alpha, \beta)] \\ & + J^\#(\beta)[g(\gamma, \alpha)] + g([\alpha, \beta], \gamma) - g([\beta, \gamma], \alpha) + g([\gamma, \alpha], \beta) \nonumber
\end{align}
where $\alpha, \beta, \gamma \in \Omega^1(M)$ and $J^\#: T^*\mathcal{Z} \rightarrow T\mathcal{Z} $ is the anchor map defined by 
\begin{align}
\left<\beta, J^\#(\alpha)\right> =  J(\alpha, \beta).
\end{align}
In coordinate form
\begin{align}
J^\#(\mathbf{d}f)[g] = \frac{\partial f}{\partial z^i} J^{ij}\frac{\partial g}{\partial z^j}.
\end{align}
The Kozul bracket on general one-forms can be obtained using the formula
\begin{align}
[\alpha, f(z)\beta] = f(z) [\alpha, \beta] + J^\#(\alpha)[f(z)]\beta.
\end{align}
It should be noted $D$ is not a covariant derivative, which is defined with tangent vectors. Even so, it can be shown $D$ satisfies similar linearity properties
\begin{align}
D_{\alpha + \beta} \gamma &= D_{\alpha} \gamma + D_{\beta} \gamma \\ 
D_{f\alpha} \gamma &= fD_\alpha \gamma \nonumber \\
D_\alpha (\beta + \gamma) &= D_\alpha\beta + D_{\alpha} \gamma \nonumber \\
D_{\alpha}(f\gamma) &= f D_{\alpha}\gamma + J^\#(\alpha)[f]\gamma. \nonumber 
\end{align}
Letting $\alpha = \mathbf{d}f$, $\beta = \mathbf{d}h$, and $\gamma = \mathbf{d}z^\delta$, we can write the contravariant derivate as 
\begin{align}
2g^{\delta a}(D_{\mathbf{d}f} \mathbf{d}h)_a  = \,&\frac{\partial f}{\partial z^i} J^{ij}\frac{\partial}{\partial z^j}\left[g^{k\delta} \frac{\partial h}{\partial z^k} \right] + J^{i \delta }\frac{\partial}{\partial z^i}\left[g^{kl} \frac{\partial f}{\partial z^k} \frac{\partial h}{\partial z^l}\right] \\  + &\frac{\partial h}{\partial z^i} J^{ij}\frac{\partial}{\partial z^j}\left[g^{k\delta }  \frac{\partial f}{\partial z^k}\right]+ g^{k\delta}\frac{\partial}{\partial z^k}\left[ J^{ij}\frac{\partial f}{\partial z^i}\frac{\partial g}{\partial z^j}\right]  \nonumber  \\& -  g^{kl}\frac{\partial}{\partial z^k}\left[ J^{i\delta}\frac{\partial h}{\partial z^i}\right]   \frac{\partial f}{\partial z^l} -  g^{kl}\frac{\partial}{\partial z^k}\left[ J^{jk }\frac{\partial f}{\partial z^j}\right]   \frac{\partial h}{\partial z^l}. \nonumber
\end{align}
It is convenient to introduce  
\begin{align}
\Gamma^{ij}_l  \colon = (D_{\mathbf{d}z^i} \mathbf{d}z^j)_k =\,
 &\frac{1}{2} g_{kl}\left[ J^{ia}\frac{\partial g^{jk}}{\partial z^a}- J^{ k a}\frac{\partial g^{ij}}{\partial z^a} + J^{ja}\frac{\partial g^{ik}}{\partial z^a}\right] \\ & +\frac{1}{2} g_{kl}\left[  g^{ka}\frac{\partial J^{ij}}{\partial z^a}   -  g^{ai}\frac{\partial J^{jk}}{\partial z^a}  -  g^{aj}\frac{\partial J^{ik}}{\partial z^a}  \right]. \nonumber
\end{align}
$D$ is the unique connection that is both torison-free
\begin{align}
D_\alpha \beta - D_\beta \alpha = [\alpha, \beta] 
\end{align}
and metric compatible 
\begin{align}
J^\# (\alpha) [g(\beta, \gamma) ] = g(D_\alpha \beta, \gamma) + g(\beta, D_\alpha \gamma).
\end{align}
In coordinates, these conditions read 
\begin{align}
\Gamma^{ij}_k - \Gamma^{ji}_k = \frac{\partial J^{ij}}{\partial z^k} 
\end{align}
and 
\begin{align}
J^{ia}\frac{\partial g^{jk}}{\partial z^j}   = g^{ka}\Gamma^{ij}_a   + g^{ja} \Gamma_a^{ik}.
\end{align}

We define the contravariant Riemann curvature tensor in a way formally reminiscent of the usual curvature tensor 
\begin{align}
R(\alpha, \beta) \gamma = D_\alpha D_\beta \gamma - D_\beta D_\alpha \gamma - D_{[\alpha, \beta]}\gamma.
\end{align}
This tensor gives us a natural 4-bracket $( \cdot,\cdot; \cdot, \cdot):  C^\infty(\mathcal{Z}) \times C^\infty(\mathcal{Z}) \times C^\infty(\mathcal{Z}) \times C^\infty(\mathcal{Z}) \rightarrow C^\infty(\mathcal{Z})$ defined by 
\begin{align}
(F,K; G, N) \colon =  g( R(\mathbf{d}F,\mathbf{d}K)\mathbf{d}G, \mathbf{d}N)
\end{align}
or in coordinates, noting the raised index, 
\begin{align}
(F,K; G, N) =   R^{ijkl}\frac{\partial F}{\partial z^i}\frac{\partial K}{\partial z^j}\frac{\partial G}{\partial z^k}\frac{\partial N}{\partial z^l}.
\end{align}
This bracket inherits the following symmetries  
\begin{align}
(F,K; G,N) = -(K,F;G,N) = -(F,K;N,G) = (G,N; F,K) 
\end{align}
in addition to the symmetries obtained by the first and second Bianchi identity
(see e.g.\cite{nakahara}). 

Given a Hamiltonian $H$, we can define a bracket analogous to the operator describing geodesic deviation
\begin{align}
(F, G) \colon =  (F,H;G,H).
\end{align}
Notice, that by the symmetries of the 4-bracket, $(F,G)$ is both symmetric and has $H$ in its kernel, precisely the necessary conditions for a metriplectic bracket. 
\section*{Free Rigid Body Bracket}
Before we use the formalism of Riemann-Poisson Geometry to add dissipation to the heavy top, we first attempt to understand the geometry of  the simpler free rigid body. Using angular momentum as phase space coordinates ($z^i = L^i$, $\mathcal{Z} =\mathbb{R}^3$), the Poisson bracket for the FRB system realizes the $\mathfrak{so}(3)$ algebra
\begin{align}
\{L^i,L^j\} = - \epsilon^{ijk}L^k.
\end{align}
The Hamiltonian for this system is 
\begin{align}
H = \frac{1}{2}\left(\frac{L^1}{I_1} + \frac{L^2}{I_2} + \frac{L^3}{I_3}   \right) .
\end{align}

To introduce dissipation to this system, we first consider a \say{dissipative metric} for the system. The simplest possible choice is the Euclidian (or equivalently the Cartan-killing) metric 
\begin{align}
g^{ij} = \delta^{ij} 
\end{align}
For a general Lie-Poisson bracket corresponding to a semi-simple, compact algebra with structure constants $C^{ij}_k$, the contravariant derivative takes a very simple form when we use the Cartan-Killing form as the metric
\begin{align}
\Gamma^{ij}_k = \frac{1}{2}C^{ij}_k .
\end{align}
The totally contravariant curvature tensor is also quite simple
\begin{align}
R^{ijk}_l = \frac{1}{4}C^{jk}_aC^{ia}_l - \frac{1}{4}C^{ik}_aC^{jal} + \frac{1}{2}C^{ij}_aC^{ka}_l.
\end{align}

For the $\mathfrak{so}(3)$ case at hand, the structure constants are given by $C^{ij}_k = -\epsilon_{ijk}$ and our curvature tensor is given by 
\begin{align}
R^{ijkl} = \frac{1}{4}(\delta^{ik}\delta^{jl}-\delta^{il}\delta^{jk} ) .
\end{align}
The metriplectic bracket is, up to a constant,
\begin{align}
(F,G) = (\omega^2 \delta_{ij} - \omega_i \omega_j) \frac{\partial F}{\partial L^i}\frac{\partial G}{\partial L^j} .
\end{align}
Amazingly, this is the bracket first constructed in \cite{original} and later considered in Matterasi and Morrison \cite{metfrb} as a way to model an energy-conserving torque driving the body to rotate about a principle axis.
\section*{Heavy Top Bracket}
Inspired by the free-rigid body, we consider the heavy top system describing a rigid body in a constant gravitational field, which is given by the body frame vector $\mathbf{\Gamma}$. The Poisson bracket is \cite{osti_594419}\cite{text}
\begin{align}
\{F,G \} = -\epsilon_{ijk} L^i \frac{\partial F}{\partial L^j}\frac{\partial G}{\partial L^k} -  \epsilon_{ijk} \Gamma^i \left( \frac{\partial F}{\partial L^j}\frac{\partial G}{\partial \Gamma^k} - \frac{\partial G}{\partial L^j}\frac{\partial F}{\partial \Gamma^k}\right). 
\end{align}
In analogy with the free rigid body system from earlier, it is only natural (and as we will see, desirable) to assume the heavy top also has a Euclidean metric 
\begin{align}
g^{ij} = \delta^{ij}.
\end{align}
Organizing the phase space coordinates $\mathbf{z}^i = (\mathbf{L}, \mathbf{\Gamma})^i$, the Lie-Poisson structure of the heavy top bracket allows us to write the bracket as 
\begin{align}
\{z^i, z^j\}= C^{ij}_k z^k
\end{align}
where $C^{ij}_k$ are the structure constants of the semi-direct product algebra $\mathfrak{so}(3) \ltimes \mathbb{R}^3$. 

For any Lie-Poisson bracket with the Euclidean metric, the connection coefficients are (a similar formula can be found in \cite{milnor})  
\begin{align}
 \Gamma^{ij}_k = \frac{1}{2}(C^{ij}_k - C^{jk}_i + C^{ki}_j)
\end{align}
and the curvature tensor is 
\begin{align}
R^{ijk}_l = & \frac{1}{4}(C^{jk}_a - C^{ka}_j + C^{aj}_k)(C^{ia}_l -  C^{al}_i + C^{li}_a)  
\\ &  
-\frac{1}{4}(C^{ik}_a - C^{ka}_i + C^{ai}_k)(C^{ja}_l - C^{al}_j + C^{lj}_a) -\frac{1}{2}C^{ij}_a(C^{ak}_l - C^{kl}_a + C^{la}_k). \nonumber
\end{align}

In general, a six-dimensional system like the heavy top will have $105$  independent components in the curvature tensor, and over a $1000$ dependant terms. Fortunately, calculations like these are easily done computationally. For the heavy top, we find that the curvature tensor is quite sparse with the nonzero terms being
\begin{align}
\frac{1}{4}&= R^{1212} = R^{1313} = R^{2121} = R^{2323} = R^{3131} = R^{3232}  \\&=-R^{2112} = -R^{3113} = -R^{1221} = -R^{3223} = -R^{1331} = -R^{2332}. \nonumber
\end{align}
This expression is even simpler than it appears. Every term involving a $\mathbf{\Gamma}$ index vanishes. Furthermore, for $i,j,k,l \in \{1,2,3\}$
\begin{align}
R^{ijkl} = \frac{1}{4}(\delta^{ik}\delta^{jl}-\delta^{il}\delta^{jk} ).
\end{align}
That is, restricted to $\mathbf{L}$ indices only, the curvature tensor is exactly that of the free rigid body.

From here on, we assume the top is symmetric ($I_1 = I_2$).The Hamiltonian for the symmetric heavy top system is given by 
\begin{align}
H = \frac{1}{2}\left(\frac{L^1}{I_1} + \frac{L^2}{I_1} + \frac{L^3}{I_3}   \right) 
+ \xi \Gamma^3 . 
\end{align}
Plugging this into the curvature tensor, the heavy top metriplectic bracket is precisely the same bracket considered in Morrison and Matterasi to describe the FRB 
\begin{align}
(F,G) = (\omega^2 \delta^{ij} - \omega^i \omega^j) \frac{\partial F}{\partial L^i}\frac{\partial G}{\partial L^j}. 
\end{align}

Unlike the free rigid body, the total angular momentum $L^2$ is no longer a Casimir invariant. Rather, there are two new Casimirs up to composition with an analytic function \cite{osti_594419}
\begin{align}
\vec{\Gamma} \cdot \vec{L} 
\end{align}
and 
\begin{align}
\vec{\Gamma}^2. 
\end{align}
The latter choice of Casimir trivializes the metriplectic dynamics, so the only suitable choice of generating function for the dissipation has the form 
\begin{align}
S = C(\vec{\Gamma} \cdot \vec{\omega} ) 
\end{align}
where $C$ is analytic. 

The dissipative equations of motion are 
\begin{align}
&\dot{f} = \{f,H\} + (f,S) =\\ &-\epsilon^{ijk} L^i \frac{\partial f}{\partial L^j}\omega^k-  \Gamma^i \left( \epsilon^{ij3} \frac{\partial f}{\partial L^j}\xi- \epsilon^{ijk}\omega^j \frac{\partial f}{\partial \Gamma^k}\right) + C'(\vec{\Gamma} \cdot \vec{\omega})(\omega^2 \delta^{ij} - \omega^i \omega^j) \frac{\partial f}{\partial L^j}\Gamma^i. \nonumber
\end{align}
We can expand this equation as a system of $6$ ODEs
\begin{align}
\dot{L}^1  &=  -L^2 L^3\left( \frac{1}{I_1}- \frac{1}{I_3} \right) +  \xi \Gamma^2   + C'(\mathbf{\Gamma} \cdot \mathbf{L}) \left(\left[ \frac{(L^2)^2}{I_1^2} +  \frac{(L^3)^2}{I_3^2}\right]\Gamma^1 -\frac{L^1 L^2 }{ I^2_1}\Gamma^2 -\frac{ L^1 L^3 }{I_1 I_3}\Gamma^3  \right )  \\
\dot{L}^2  &=  L^1 L^3\left( \frac{1}{I_1}- \frac{1}{I_3} \right) -  \xi \Gamma^1  + C'(\mathbf{\Gamma} \cdot \mathbf{L})\left(-\frac{L^1 L^2 }{ I^2_1}\Gamma^1 + \left[ \frac{(L^1)^2}{I_1^2} +  \frac{(L^3)^2}{I_3^2}\right]\Gamma^2  -\frac{ L^2 L^3 }{I_1 I_3}\Gamma^3  \right ) \nonumber \\
\dot{L}^3  &=   C'(\mathbf{\Gamma} \cdot \mathbf{L})\left( -\frac{L^1 L^3 }{ I_1I_3 }\Gamma^1 -\frac{ L^2 L^3 }{I_1 I_3}\Gamma^2 + \left[ \frac{(L^2)^2}{I_1^2} +  \frac{(L^1)^2}{I_1^2}\right]\Gamma^3 \right ) \nonumber \\
\dot{\Gamma}^1 &=    \Gamma^{2}\frac{L^3}{I_3} - \Gamma^{3}\frac{L^2}{I_1} \nonumber\\
\dot{\Gamma}^2 &=   \Gamma^{3}\frac{L^1}{I_1}-\Gamma^{1}\frac{L^3}{I_3} \nonumber\\
\dot{\Gamma}^3 &=   \Gamma^{1} \frac{L^2}{I_1} - \Gamma^{2} \frac{L^1}{I_1}. \nonumber
\end{align}
Notice that the dissipation does not affect the equation of motion for $\mathbf{\Gamma}$, as we should expect from any physically realistic system. In fact, had we naively used a bracket projecting out the Hamiltonian, this would no longer be true. It is worthwhile to note that $S$ is increased by the dynamics since 
 \begin{align}
 \dot{S} =  (C'(\vec{\Gamma} \cdot \vec{\omega}))^2 \Gamma^2\omega^2 \sin^2\theta \geq 0 
 \end{align}
 where $\theta$ measures the angle between $\vec{\omega}$ and $\vec{\Gamma}$. 
\section*{Dynamics of The Heavy Top}
The first analytic function we try is $C(x) = \lambda x$. The equations of motion are 
  \begin{align}
\dot{L}^1  &=  -L^2 L^3\left( \frac{1}{I_1}- \frac{1}{I_3} \right) +  \xi \Gamma^2   + \lambda \left(\left[ \frac{(L^2)^2}{I_1^2} +  \frac{(L^3)^2}{I_3^2}\right]\Gamma^1 -\frac{L^1 L^2 }{ I^2_1}\Gamma^2 -\frac{ L^1 L^3 }{I_1 I_3}\Gamma^3  \right )  \\
\dot{L}^2  &=  L^1 L^3\left( \frac{1}{I_1}- \frac{1}{I_3} \right) -  \xi \Gamma^1  + \lambda\left(-\frac{L^1 L^2 }{ I^2_1}\Gamma^1 + \left[ \frac{(L^1)^2}{I_1^2} +  \frac{(L^3)^2}{I_3^2}\right]\Gamma^2  -\frac{ L^2 L^3 }{I_1 I_3}\Gamma^3  \right ) \nonumber \\
\dot{L}^3  &=  \lambda \left( -\frac{L^1 L^3 }{ I_1I_3 }\Gamma^1 -\frac{ L^2 L^3 }{I_1 I_3}\Gamma^2 + \left[ \frac{(L^2)^2}{I_1^2} +  \frac{(L^1)^2}{I_1^2}\right]\Gamma^3 \right ) \nonumber \\
\dot{\Gamma}^1 &=    \Gamma^{2}\frac{L^3}{I_3} - \Gamma^{3}\frac{L^2}{I_1} \nonumber\\
\dot{\Gamma}^2 &=   \Gamma^{3}\frac{L^1}{I_1}-\Gamma^{1}\frac{L^3}{I_3} \nonumber\\
\dot{\Gamma}^3 &=   \Gamma^{1} \frac{L^2}{I_1} - \Gamma^{2} \frac{L^1}{I_1}. \nonumber
\end{align}
 It's not hard to see that the metriplectic heavy top has an equilibrium when $\Gamma^{1,2} = 0$ and $L^{1,2} = 0$. Given an initial configuration, we define $L^3_*>0$ and $\Gamma_*^3>0$ to be the physically realizable equilibrium consistent with $\Gamma^{1,2} = 0$ and $L^{1,2} = 0$.  Linearizing the equations of motion around this equilibrium, we get the equations 
 \begin{align}
\dot{\delta L}^1  &=  - L_*^3\left( \frac{1}{I_1}- \frac{1}{I_3} \right)\delta L^2 +  \xi \delta\Gamma^2   + \lambda \left(\frac{(L_*^3)^2}{I_3^2}\delta\Gamma^1  -\frac{  L_*^3\Gamma_*^3 }{I_1 I_3} \delta L^1  \right ) \\
\dot{\delta L}^2  &=  L_*^3\left( \frac{1}{I_1}- \frac{1}{I_3} \right) \delta L^1 -  \xi \delta\Gamma^1  + \lambda \left( \frac{(L_*^3)^2}{I_3^2}\delta\Gamma^2  -\frac{L_*^3 \Gamma_*^3}{I_1 I_3}  \delta L^2  \right )\nonumber\\
\dot{\delta L}^3  &=  0 \nonumber\\
\dot{\delta \Gamma}^1 &=   \frac{L_*^3}{I_3}  \delta \Gamma^{2} - \frac{\Gamma_*^{3}}{I_1}\delta L^2 \nonumber\\ 
\dot{\delta \Gamma}^2 &=   \frac{\Gamma_*^{3}}{I_1}\delta L^1-\frac{L_*^3}{I_3}\delta \Gamma^{1} \nonumber \\
\dot{\delta \Gamma}^3 &=  0. \nonumber  
\end{align}
We can express these equations compactly as
\begin{align}
\dot{\delta\mathbf{z}} = M\delta\mathbf{z}.
\end{align}

The general solution for the spectrum of $M$ isn't particularly enlightening beyond the fact that the eigenvalues are always of the form $(0,0,A,A^*,B,B^*)$. Depending on the initial conditions, either (a.) $Re(A)$ and $Re(B)$ differ in sign or (b.) both $Re(A)<0$ and $Re(B)<0$. The latter case is of particular interest since it implies the system is linearly stable. If we impose that $I_3 = 2I_1$, the spectrum of $M$ is a lot easier to understand with 
\[
A = \frac{-\Gamma_*^3L_*^3\lambda - \sqrt{-4(I_1)^2 (L_*^3)^2 +(\Gamma_*^3)^2(L_*^3)^2\lambda^2 + 16\Gamma_*^3(I_1)^3\xi }}{4(I_1)^2}
\] 
\[
B = \frac{-\Gamma_*^3L_*^3\lambda + \sqrt{-4(I_1)^2 (L_*^3)^2 +(\Gamma_*^3)^2(L_*^3)^2\lambda^2 + 16\Gamma_*^3(I_1)^3\xi }}{4(I_1)^2}.
\] 
When the root is real, the system is linearly stable if and only if 
\[
 (L_*^3)^2 \geq  4\Gamma_*^3I_1\xi .
\]
When the root is complex, 
the system is always linearly stable, with the dynamics oscillating towards the equilibrium $L^{1,2} = \Gamma^{1,2} = 0$.
\\

Computationally modeling the dynamics in Mathematica, it seems the relaxation behavior of this system can be well understood in terms of its linearization, provided the top does not fall. For example, in arbitrary units, we can let $I_1 = 1 = \frac{1}{2}I_3$, $\lambda = 0.1$, and $\xi = 1$. If we start with with initial conditions such that $L_*^3 \approx 5$, $\Gamma_*^3 \approx 3$ then
\begin{align}
\text{Eigenvalues}(M) \approx (0,0, -.38 + 1.76i , -.38 - 1.76i, -.38 + 1.76i, -.38 - 1.76i). 
\end{align}
\begin{figure}[h]
    \centering
    \subfloat[\centering $L^1$ vs. $t$  ]{{\includegraphics[scale=.4]{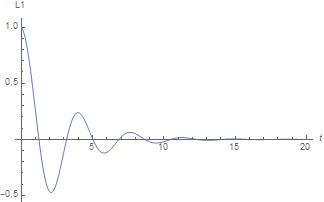} }}%
    \subfloat[\centering $L^2$ vs. $t$ ]{{\includegraphics[scale=.4]{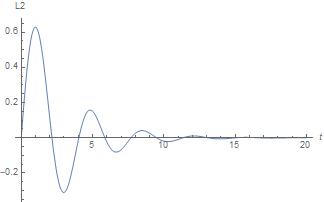} }}%
    \\
    \subfloat[\centering $L^3$ vs. $t$ ]{{\includegraphics[scale=.4]{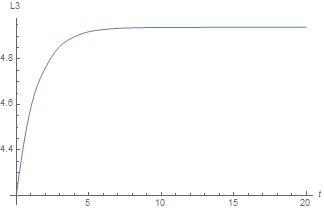} }}%
    \subfloat[\centering $\Gamma^1$ vs. $t$]{{\includegraphics[scale=.4]{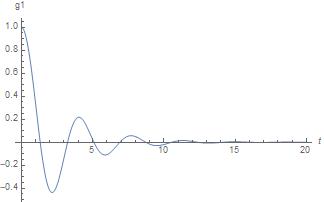} }}%
    \\
    \subfloat[\centering $\Gamma^2$ vs. $t$]{{\includegraphics[scale=.4]{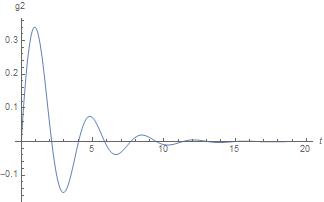} }}%
    \subfloat[\centering $\Gamma^3$ vs. $t$]{{\includegraphics[scale=.4]{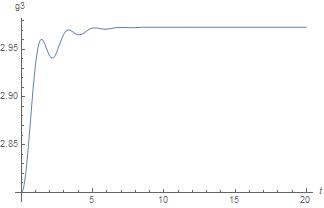} }}%
    \caption{$\lambda = 0.1; C(x) = \lambda x$ }%
    \label{fig:example1}%
\end{figure}
The linear dynamics predict that the metriplectic heavy top solutions will decay towards equilibrium while oscillating. Modeling the full nonlinear dynamics with initial conditions $\mathbf{\Gamma}(t=0) = (1,0,2.8)$ and $\mathbf{L}(t=0) = (1,0,4.2)$, we see that the behavior of the system qualitatively approximates the linear dynamics, even with large perturbations from equilibrium (figure 1). This is an interesting point, since $S$ is not a valid Lyapunov function. Likely, this stability comes from a sort of constrained optimization on the surfaces of constant energy.

If we instead fix $\lambda =1$ then
\begin{align}
\text{Eigenvalues}(M) \approx (0,0, -7.03 , -7.03, -.46, -.46),
\end{align}
which is again stable, but this time we see no oscillatory behavior in the linearized dynamics. Changing our initial conditions to $\mathbf{\Gamma}(t=0) = (.3,0,3)^T$ and $\mathbf{L}(t=0) = (.5,0,5.2)$ so the top doesn't fall, we again observe the linear dynamics is qualitatively similar to the relaxation behavior of the full nonlinear dynamics (figure 2).
\begin{figure}[h]
    \centering
    \subfloat[\centering $L^1$ vs. $t$  ]{{\includegraphics[scale=.4]{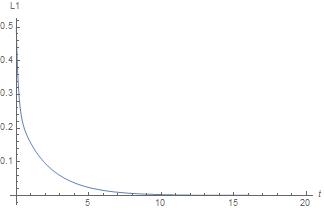} }}%
    \subfloat[\centering $L^2$ vs. $t$ ]{{\includegraphics[scale=.4]{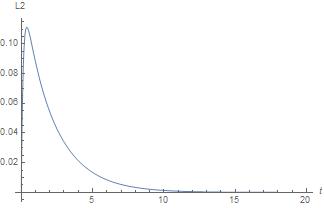} }}%
    \\
    \subfloat[\centering $L^3$ vs. $t$ ]{{\includegraphics[scale=.4]{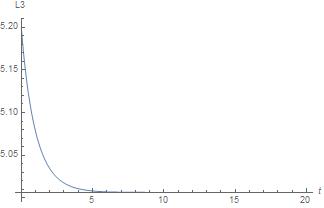} }}%
    \subfloat[\centering $\Gamma^1$ vs. $t$]{{\includegraphics[scale=.4]{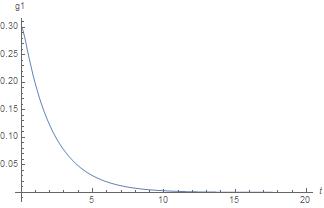} }}%
    \\
    \subfloat[\centering $\Gamma^2$ vs. $t$ ]{{\includegraphics[scale=.4]{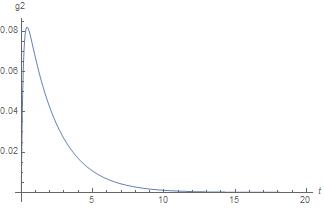} }}%
    \subfloat[\centering $\Gamma^3$ vs. $t$]{{\includegraphics[scale=.4]{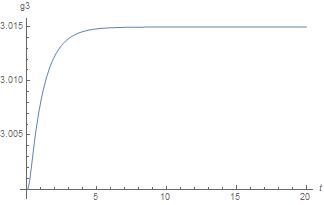} }}%
    \caption{$\lambda = 1; C(x) = \lambda x$ }%
    \label{fig:example1}%
\end{figure} 

Provided we change the phase space from $\mathbb{R}^6$ to  $\mathbb{R}^6-S$, where $S$ is the vanishing set of $\mathbf{\Gamma} \cdot \mathbf{L}$, we can also consider the dynamics of the heavy top with an otherwise singular generating function. In particular, we explore when $S = \lambda \log(\mathbf{\Gamma} \cdot \mathbf{L}) $. The equations of motion are 
\begin{align}
 \dot{L}^1  &=  -L^2 L^3\left( \frac{1}{I_1}- \frac{1}{I_3} \right) +  \xi \Gamma^2   + \frac{\lambda}{\mathbf{\Gamma} \cdot \mathbf{L}} \left(\left[ \frac{(L^2)^2}{I_1^2} +  \frac{(L^3)^2}{I_3^2}\right]\Gamma^1 -\frac{L^1 L^2 }{ I^2_1}\Gamma^2 -\frac{ L^1 L^3 }{I_1 I_3}\Gamma^3  \right )  \\
\dot{L}^2  &=  L^1 L^3\left( \frac{1}{I_1}- \frac{1}{I_3} \right) -  \xi \Gamma^1  + \frac{\lambda}{\mathbf{\Gamma} \cdot \mathbf{L}}\left(-\frac{L^1 L^2 }{ I^2_1}\Gamma^1 + \left[ \frac{(L^1)^2}{I_1^2} +  \frac{(L^3)^2}{I_3^2}\right]\Gamma^2  -\frac{ L^2 L^3 }{I_1 I_3}\Gamma^3  \right ) \nonumber \\
\dot{L}^3  &=   \frac{\lambda}{\mathbf{\Gamma} \cdot \mathbf{L}} \left( -\frac{L^1 L^3 }{ I_1I_3 }\Gamma^1 -\frac{ L^2 L^3 }{I_1 I_3}\Gamma^2 + \left[ \frac{(L^2)^2}{I_1^2} +  \frac{(L^1)^2}{I_1^2}\right]\Gamma^3 \right ) \nonumber \\
\dot{\Gamma}^1 &=    \Gamma^{2}\frac{L^3}{I_3} - \Gamma^{3}\frac{L^2}{I_1} \nonumber\\
\dot{\Gamma}^2 &=   \Gamma^{3}\frac{L^1}{I_1}-\Gamma^{1}\frac{L^3}{I_3} \nonumber\\
\dot{\Gamma}^3 &=   \Gamma^{1} \frac{L^2}{I_1} - \Gamma^{2} \frac{L^1}{I_1}. \nonumber
\end{align}
Again, this system has an equilibrium $\Gamma^{1,2} = 0$ and $L^{1,2} = 0$. Letting be $L^3_*>0$ and $\Gamma_*^3>0$ be the realizable equilibrium conditions, the linearized equations are
 \begin{align}
\dot{\delta L}^1  &=  - L_*^3\left( \frac{1}{I_1}- \frac{1}{I_3} \right)\delta L^2 +  \xi \delta\Gamma^2   + \frac{\lambda}{L_*^3\Gamma_*^3}\left(\frac{(L_*^3)^2}{I_3^2}\delta\Gamma^1  -\frac{  L_*^3\Gamma_*^3 }{I_1 I_3} \delta L^1  \right ) \\
\dot{\delta L}^2  &=  L_*^3\left( \frac{1}{I_1}- \frac{1}{I_3} \right) \delta L^1 -  \xi \delta\Gamma^1  + \frac{\lambda}{L_*^3\Gamma_*^3} \left( \frac{(L_*^3)^2}{I_3^2}\delta\Gamma^2  -\frac{L_*^3 \Gamma_*^3}{I_1 I_3}  \delta L^2  \right ) \nonumber\\
\dot{ \delta L}^3  &=  0 \nonumber\\
\dot{ \delta\Gamma}^1 &=   \frac{L_*^3}{I_3}  \delta \Gamma^{2} - \frac{\Gamma_*^{3}}{I_1}\delta L^2 \nonumber\\
\dot{ \delta\Gamma}^2 &=   \frac{\Gamma_*^{3}}{I_1}\delta L^1-\frac{L_*^3}{I_3}\delta \Gamma^{1} \nonumber\\
\dot{ \delta \Gamma}^3 &=  0.\nonumber 
\end{align}  
Again, we may write 
\[
\delta\dot{\mathbf{z}} = M \delta \mathbf{z}.
\]
Provided $2I_1 = I_3$, the eigenvalues of $M$ are $(0,0,A,A^*,B,B^*)$ where 
\[
A =  \frac{-\Gamma_*^3\lambda - \sqrt{-4(\Gamma_*^3)^2(I_1)^2 (L_*^3)^2 +(\Gamma_*^3)^2\lambda^2 + 16\Gamma_*^3(I_1)^3\xi }}{4(I_1)^2\Gamma_*^3}
\]
\[
B = \frac{-\Gamma_*^3\lambda + \sqrt{-4(\Gamma_*^3)^2(I_1)^2 (L_*^3)^2 +(\Gamma_*^3)^2\lambda^2 + 16\Gamma_*^3(I_1)^3\xi }}{4(I_1)^2\Gamma_*^3}.
\]
\begin{figure}[h]
    \centering
    \subfloat[\centering $L^1$ vs. $t$  ]{{\includegraphics[scale=.4]{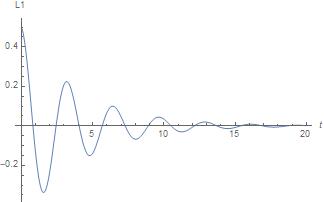} }}%
    \subfloat[\centering $L^2$ vs. $t$ ]{{\includegraphics[scale=.4]{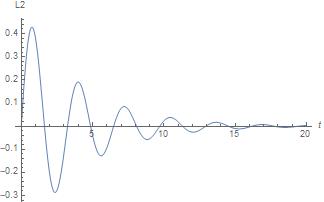} }}%
    \\
    \subfloat[\centering $L^3$ vs. $t$ ]{{\includegraphics[scale=.4]{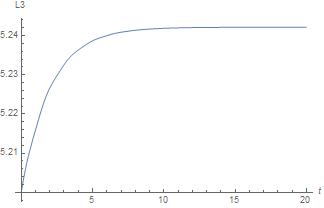} }}%
    \subfloat[\centering $\Gamma^1$ vs. $t$]{{\includegraphics[scale=.4]{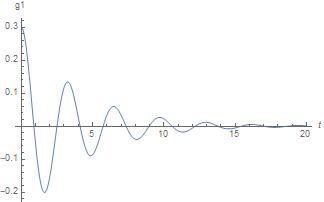} }}%
    \\
    \subfloat[\centering $\Gamma^2$ vs. $t$ ]{{\includegraphics[scale=.4]{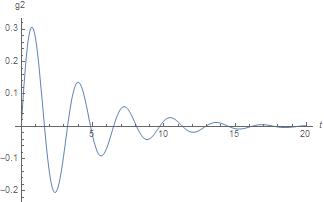} }}%
    \subfloat[\centering $\Gamma^3$ vs. $t$]{{\includegraphics[scale=.4]{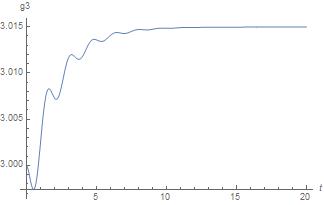} }}%
    \caption{$\lambda = 1; C(x) = \lambda \log(x) $ }%
    \label{fig:example1}%
\end{figure}
The system is linearly stable when 
\[
\Gamma_*^3 (L_*^3)^2>  4  I_1\xi 
\]

For the sake of comparison to the $C(x) = \lambda x$ case, we use $I_1 =1$, $I_1 = 2$, $\xi = 1$, and $\lambda =1$ with the initial conditions $\Gamma = (.3,0,3)$ and $L = (.5, 0, 5.2)$. We see the nonlinear system stable relaxes to $\Gamma^3_* \approx 3.0$ and $L^3_* \approx 5.2$ (figure 3). This is reflected in the matrix for the linear equations of motion, which has eigenvalues 
\[
\text{Eigenvalues}(M) \approx (0,0, -0.25 + 2i, -0.25 + 2i,-0.25 + 2i,-0.25 + 2i).
\]
Also for the sake of comparison, we can also choose $\lambda = .1$, $\mathbf{L}(t=0) = (1,0,4.2)$, and $\mathbf{\Gamma}(t=0) = (1,0,2.8)$. We see the system relaxes to equilibrium, but at about a tenth the speed as before, agreeing with the dynamics predicted by the eigenvalues of $M$ (figure 4)
\[
\text{Eigenvalue}(M) \approx (0,0, -0.025 + 1.67i, -0.025 + 1.67i, -0.025 - 1.67i, -0.025 - 1.67i).
\]
\begin{figure}[h]
    \centering
    \subfloat[\centering $L^1$ vs. $t$  ]{{\includegraphics[scale=.4]{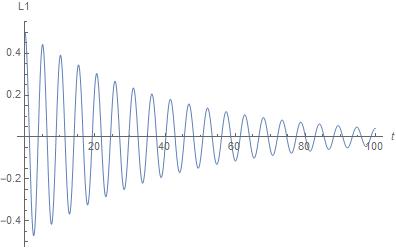} }}%
    \subfloat[\centering $L^2$ vs. $t$ ]{{\includegraphics[scale=.4]{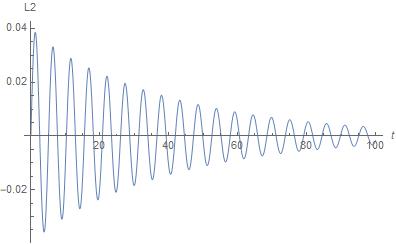} }}%
    \\
    \subfloat[\centering $L^3$ vs. $t$ ]{{\includegraphics[scale=.4]{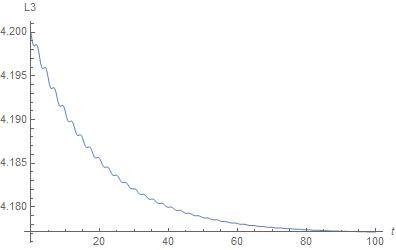} }}%
    \subfloat[\centering $\Gamma^1$ vs. $t$]{{\includegraphics[scale=.4]{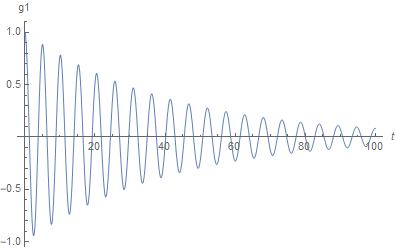} }}%
    \\
    \subfloat[\centering $\Gamma^2$ vs. $t$ ]{{\includegraphics[scale=.4]{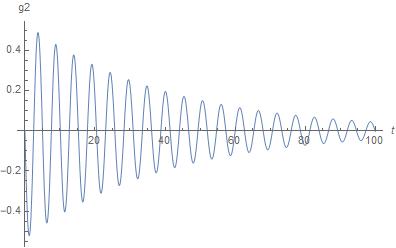} }}%
    \subfloat[\centering $\Gamma^3$ vs. $t$]{{\includegraphics[scale=.4]{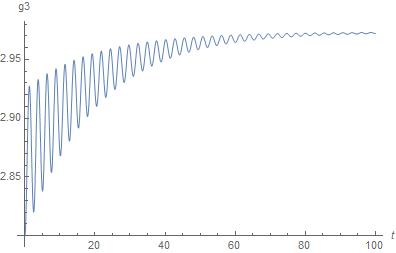} }}%
    \caption{$\lambda = 0.1; C(x) = \lambda \log(x)$ }%
    \label{fig:example1}%
\end{figure}

As a final example, we consider the case when $C(x) = \frac{\lambda}{2}x^2 $. The equations of motion are  
 \begin{align}
\dot{L}^1  &=  -L^2 L^3\left( \frac{1}{I_1}- \frac{1}{I_3} \right) +  \xi \Gamma^2   + \lambda \mathbf{\Gamma} \cdot \mathbf{L}  \left(\left[ \frac{(L^2)^2}{I_1^2} +  \frac{(L^3)^2}{I_3^2}\right]\Gamma^1 -\frac{L^1 L^2 }{ I^2_1}\Gamma^2 -\frac{ L^1 L^3 }{I_1 I_3}\Gamma^3  \right )  \\
\dot{L}^2  &=  L^1 L^3\left( \frac{1}{I_1}- \frac{1}{I_3} \right) -  \xi \Gamma^1  + \lambda \mathbf{\Gamma} \cdot \mathbf{L} \left(-\frac{L^1 L^2 }{ I^2_1}\Gamma^1 + \left[ \frac{(L^1)^2}{I_1^2} +  \frac{(L^3)^2}{I_3^2}\right]\Gamma^2  -\frac{ L^2 L^3 }{I_1 I_3}\Gamma^3  \right ) \nonumber \\
\dot{L}^3  &=   \lambda \mathbf{\Gamma} \cdot \mathbf{L}  \left( -\frac{L^1 L^3 }{ I_1I_3 }\Gamma^1 -\frac{ L^2 L^3 }{I_1 I_3}\Gamma^2 + \left[ \frac{(L^2)^2}{I_1^2} +  \frac{(L^1)^2}{I_1^2}\right]\Gamma^3 \right ) \nonumber \\
\dot{\Gamma}^1 &=    \Gamma^{2}\frac{L^3}{I_3} - \Gamma^{3}\frac{L^2}{I_1} \nonumber\\
\dot{\Gamma}^2 &=   \Gamma^{3}\frac{L^1}{I_1}-\Gamma^{1}\frac{L^3}{I_3} \nonumber\\
\dot{\Gamma}^3 &=   \Gamma^{1} \frac{L^2}{I_1} - \Gamma^{2} \frac{L^1}{I_1}, \nonumber
\end{align}
which we linearize 
 \begin{align}
\dot{\delta L}^1  &=  - L_*^3\left( \frac{1}{I_1}- \frac{1}{I_3} \right)\delta L^2 +  \xi \delta\Gamma^2   + \lambda L_*^3\Gamma_*^3\left(\frac{(L_*^3)^2}{I_3^2}\delta\Gamma^1  -\frac{  L_*^3\Gamma_*^3 }{I_1 I_3} \delta L^1  \right ) \\
\dot{\delta L}^2  &=  L_*^3\left( \frac{1}{I_1}- \frac{1}{I_3} \right) \delta L^1 -  \xi \delta\Gamma^1  + \lambda L_*^3\Gamma_*^3 \left( \frac{(L_*^3)^2}{I_3^2}\delta\Gamma^2  -\frac{L_*^3 \Gamma_*^3}{I_1 I_3}  \delta L^2  \right ) \nonumber\\
\dot{\delta L}^3  &=  0 \nonumber\\
\dot{\delta \Gamma}^1 &=   \frac{L_*^3}{I_3}  \delta \Gamma^{2} - \frac{\Gamma_*^{3}}{I_1}\delta L^2 \nonumber\\
\dot{\delta \Gamma}^2 &=   \frac{\Gamma_*^{3}}{I_1}\delta L^1-\frac{L_*^3}{I_3}\delta \Gamma^{1} \nonumber\\
\dot{\delta \Gamma}^3 &=  0. \nonumber 
\end{align}
Even with the usual caveat that $I_3 = 2I_2$, the eigenvalues of $M$ do not have a particularly nice expression. We again let $\xi = 1, I_1 = 1$, $I_3 = 2$, and $\lambda = .1$. With the initial conditions $\mathbf{\Gamma}(t=0) = (1,0,2.8)$ and $\mathbf{L}(t=0) = (1,0,4.2)$ the system is linearly stable with
\[
\text{Eigenvalues}(M) \approx (-8.31, -8.31, -0.21, -0.21, 0, 0),
\]
as reflected by the nonlinear dynamics (figure 5). 

It should always be noted, including in all the prior examples, that this system is not stable for all initial conditions, as letting $\mathbf{L}(t=0) = (1,0,2.8)^T$ and $\mathbf{\Gamma}(t=0) = (1,0,4.2)$ exemplifies (figure 6).
\begin{figure}[h]
    \centering
    \subfloat[\centering $L^1$ vs. $t$  ]{{\includegraphics[scale=.4]{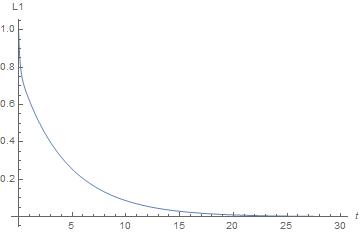} }}%
    \subfloat[\centering $L^2$ vs. $t$ ]{{\includegraphics[scale=.4]{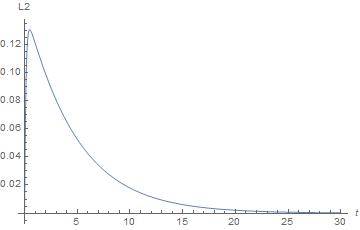} }}%
    \\
    \subfloat[\centering $L^3$ vs. $t$ ]{{\includegraphics[scale=.4]{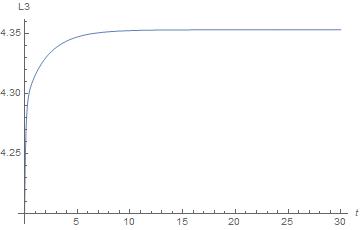} }}%
    \subfloat[\centering $\Gamma^1$ vs. $t$]{{\includegraphics[scale=.4]{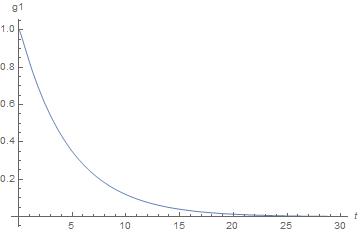} }}%
    \\
    \subfloat[\centering $\Gamma^2$ vs. $t$ ]{{\includegraphics[scale=.4]{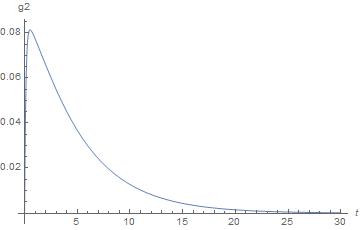} }}%
    \subfloat[\centering $\Gamma^3$ vs. $t$]{{\includegraphics[scale=.4]{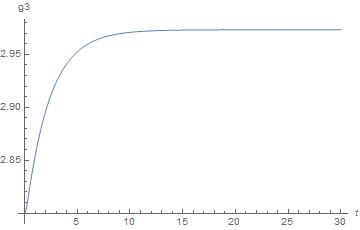} }}%
    \caption{$\lambda = 0.1$; $C(x) = \frac{\lambda}{2}x^2$; $\mathbf{\Gamma}(t=0) = (1,0,2.8)$;
 $\mathbf{L}(t=0) = (1,0,4.2)$}%
    \label{fig:example1}%
\end{figure}
\begin{figure}[H]
    \centering
    \subfloat[\centering $L^1$ vs. $t$  ]{{\includegraphics[scale=.4]{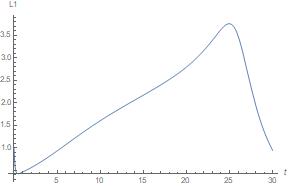} }}%
    \subfloat[\centering $L^2$ vs. $t$ ]{{\includegraphics[scale=.4]{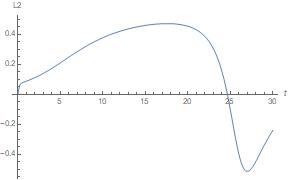} }}%
    \\
    \subfloat[\centering $L^3$ vs. $t$ ]{{\includegraphics[scale=.4]{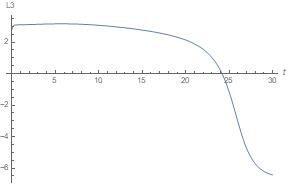} }}%
    \subfloat[\centering $\Gamma^1$ vs. $t$]{{\includegraphics[scale=.4]{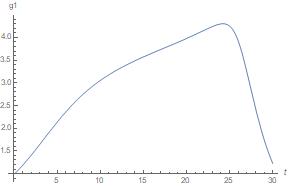} }}%
    \\
    \subfloat[\centering $\Gamma^2$ vs. $t$ ]{{\includegraphics[scale=.4]{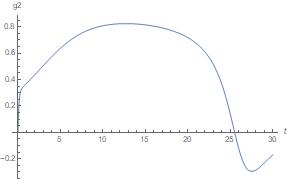} }}%
    \subfloat[\centering $\Gamma^3$ vs. $t$]{{\includegraphics[scale=.4]{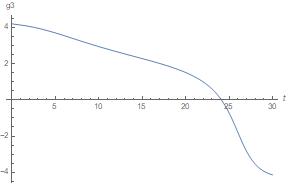} }}%
    \caption{$\lambda = 0.1$; $C(x) = \frac{\lambda}{2}x^2$; $\mathbf{L}(t=0) = (1,0,2.8)$;
 $\mathbf{\Gamma}(t=0) = (1,0,4.2)$}%
    \label{fig:example1}%
\end{figure}
Keeping with tradition, we also try $\lambda =1$ with the initial conditions $\Gamma = (.3,0,3)$ and $L = (.5, 0, 5.2)$. Like all the other systems, this initial condition is linearly stable with the nonlinear dynamics following suit (figure 7)
\[
\text{Eigevalues}(M) \approx (-11.85, -11.85, -0.317, -0.317, 0, 0).
\]
\begin{figure}[h]
    \centering
    \subfloat[\centering $L^1$ vs. $t$  ]{{\includegraphics[scale=.4]{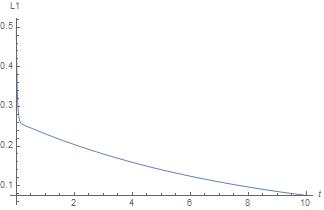} }}%
    \subfloat[\centering $L^2$ vs. $t$ ]{{\includegraphics[scale=.4]{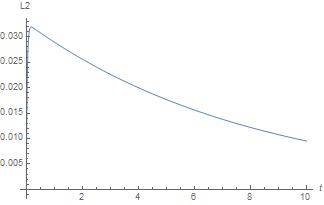} }}%
    \\
    \subfloat[\centering $L^3$ vs. $t$ ]{{\includegraphics[scale=.4]{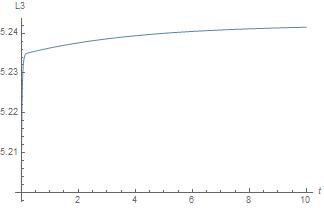} }}%
    \subfloat[\centering $\Gamma^1$ vs. $t$]{{\includegraphics[scale=.4]{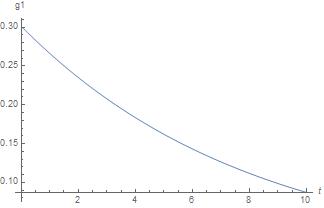} }}%
    \\
    \subfloat[\centering $\Gamma^2$ vs. $t$ ]{{\includegraphics[scale=.4]{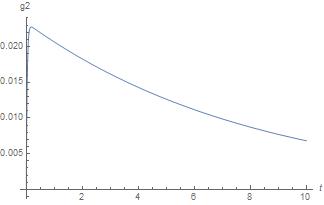} }}%
    \subfloat[\centering $\Gamma^3$ vs. $t$]{{\includegraphics[scale=.4]{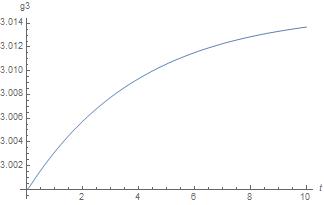} }}%
    \caption{$\lambda = 1$; $C(x) = \frac{\lambda}{2}x^2$}%
    \label{fig:example1}%
\end{figure}
\section*{Conclusion}
In this paper, we showed how the formalism of Riemann-Poisson geometry can be used to create dissipative dynamics. In particular, by considering the simplest dissipative metric on the heavy top phase space, we constructed a toy-model system that often relaxed to a stable equilibrium spinning about a principle axis, extending the work by Matterasi and Morrison \cite{metfrb} to the case where gravity cannot be neglected. We then computationally modeled the dynamics of a few systems, which while possessing the same metriplectic bracket, had their dynamics generated by different functions. 

Beyond serving as a useful toy model and proof-of-concept for more sophisticated work, the family of dynamical systems constructed in this paper have an obvious control-theoretic utility. Without dissipating any energy, beyond that required to monitor the orientation and rotation rate of a heavy top, we showed how a torque can be applied in such a way as to align a symmetric spinning body with its third moment of inertia. Perhaps more importantly, this work highlights how the language of geometry can be fitted to metriplectic dissipation, allowing for a wide class of geometric constructions to be carried into the theory of dissipative systems.

This work can be extended in a number of ways. For one, it stands open to exploration as to how different metrics can affect the allowed dynamics of this and other systems. Another, perhaps harder question, is applying this geometric formalism to field theories such as the Navier-Stokes equations. In this paper, we also left many questions about the metriplectic heavy top unanswered. Most notably, we did not fully address many interesting questions related to the stability of the system, such as what regions of the phase space are nonlinearly asymptotically stable. 
\nocite{*}
\printbibliography
\end{document}